\documentclass[prb,aps,twocolumn,showpacs,superscriptaddress]{revtex4-2}
\usepackage[utf8x]{inputenc}
\usepackage{amssymb,amsmath}
\usepackage{multirow}
\usepackage[english]{babel}
\usepackage{graphicx}
\usepackage{float}
\usepackage{wrapfig}
\usepackage{dcolumn}
\usepackage{bm}
\usepackage{hyperref}
\usepackage{natbib}
\usepackage{subfigure}
\usepackage{xcolor}
\usepackage{soul}
\usepackage[normalem]{ulem} 

\newcommand{\eg}{\emph{e.g.}, }

\newcommand{\ket}[1]{\left| #1 \right\rangle}

\newcommand{\abs}[1]{\left| #1 \right|}

\newcommand{\dg}{{^{\dagger}}}

\newcommand{\op}[1]{\hat{#1}}

\begin{document}

\title{Optimal control of a cavity-mediated iSWAP gate between silicon spin qubits}
\author{Steve Young}
\affiliation{Center for Computing Research, Sandia National Laboratories, Albuquerque, New Mexico, USA}
\author{N. Tobias Jacobson}
\affiliation{Center for Computing Research, Sandia National Laboratories, Albuquerque, New Mexico, USA}
\author{Jason R. Petta}
\affiliation{Department of Physics, Princeton University, Princeton, New Jersey 08544, USA}

\begin{abstract}
Semiconductor spin qubits may be coupled through a superconducting cavity to generate an entangling two-qubit gate. However, the fidelity of such an operation will be reduced by a variety of error mechanisms such as charge and magnetic noise, phonons, cavity loss, transitions to non-qubit states and, for electrons in silicon, excitation into other valley eigenstates. Here, we model the effects of these error sources and the valley degree of freedom on the performance of a cavity-mediated two-qubit iSWAP gate. For valley splittings inadequately large relative to the interdot tunnel coupling within each qubit, we find that valley excitation may be a limiter to the fidelity of this two-qubit gate. In addition, we show tradeoffs between gating times and exposure to various error sources, identifying optimal operating regimes and device improvements that would have the greatest impact on the fidelity of the cavity-mediated spin iSWAP. Importantly, we find that while the impact of charge noise and phonon relaxation favor operation in the regime where the qubits are most spin-like to reduce sensitivity to these sources of noise, the combination of hyperfine noise and valley physics  shifts the optimal regime to charge-like qubits with stronger effective  spin-photon coupling so that gate times can be made as short as possible.  In this regime, the primary limitation is the need to avoid Landau-Zener transitions as the gate is implemented.
\end{abstract}
	
\maketitle

\section{Introduction}
\label{sec:Introduction}
Electron spins in electrostatically-defined quantum dots continue to show promise as a platform for quantum information processing. Such devices make use of established materials and fabrication processes, and have the desirable property of an intrinsically compact ($\sim$ tens of nm) qubit size. High fidelity single- and two-qubit gates have been demonstrated \cite{Yoneda2018,Yang2019,Huang2019,Xue2019,Noiri2022,Ha2021,Mills2021,Xue2022}, with two-qubit interactions routinely realized through the modulation of the Heisenberg exchange coupling between electron spins in neighboring dots \cite{Petta2005,Sigillito2019,Takeda2021}. However, for the purpose of realizing a longer-range quantum bus it may be desirable to couple electron spin qubits over a larger length scale than the tens of nm separation between dots \cite{Vandersypen2017}.

A candidate approach to realizing long-distance spin-spin coupling is to employ a high quality factor superconducting cavity as an intermediary between spin qubits \cite{Samkharadze2018,Landig2018,Mi2018,Benito2019,Borjans2020,Borjans2020b}. The cavity may enable entangling interactions over a length scale exceeding a centimeter, many orders of magnitude more distant than is realizable via direct exchange coupling. Through the use of micromagnets to engineer a significant spin-orbit coupling \cite{Kawakami2014}, spin-spin coupling may be facilitated by tuning both spin qubits into a regime where the spin and charge degrees of freedom are coupled \cite{Petersson2012,Viennot2015,Stockklauser2017,Samkharadze2018,vanWoerkom2018,Harvey-Collard2020,HarveyCollard2022}.

Cavity-mediated spin-spin interaction requires transduction between spin and orbital degrees of freedom and the participation of an additional quantum mechanical degree of freedom, the cavity mode. Spin-photon coupling has been demonstrated for Loss-DiVincenzo (LD) \cite{Mi2018,Samkharadze2018} and resonant exchange qubits \cite{Landig2018}, resonant spin-spin coupling has been  achieved \cite{Borjans2020}, and indications of dispersive spin-spin coupling have been reported with LD spin qubits \cite{HarveyCollard2022}. Here, we quantitatively investigate the fidelity-limiting mechanisms for such a two-qubit gate.

In this paper we model the two-qubit interaction between a pair of electron spins, each occupying a double quantum dot (DQD) and coupled to a common cavity mode. We consider the full, open-system dynamics of this system operating as a quantum gate.  Recent theoretical work investigating the performance of spin-cavity \cite{Benito2017} and cavity-mediated spin-spin coupling \cite{Benito2019,Warren2019} has made several assumptions that we relax, allowing us to address important questions about practical performance limits. For one, it has been assumed that operation occurs in a regime where the intra-cavity photon number $n_c$ $\ll$ 1 and its degrees of freedom can be eliminated via transformation. We explicitly account for the cavity mode itself, avoiding the assumption of dispersive coupling where the cavity is only virtually populated \cite{Benito2019,Warren2019}. As a result, we are able to probe a continuum of coupling regimes from dispersive to resonant, allowing us to explore the relative influence of various error mechanisms. Importantly, we are able to directly study the impact of the control sequence on fidelity, including an assessment of non-adiabatic effects induced by the control modulation that result in leakage into cavity modes.

\begin{figure*}[th]
	\centering
	\includegraphics[width=\textwidth]{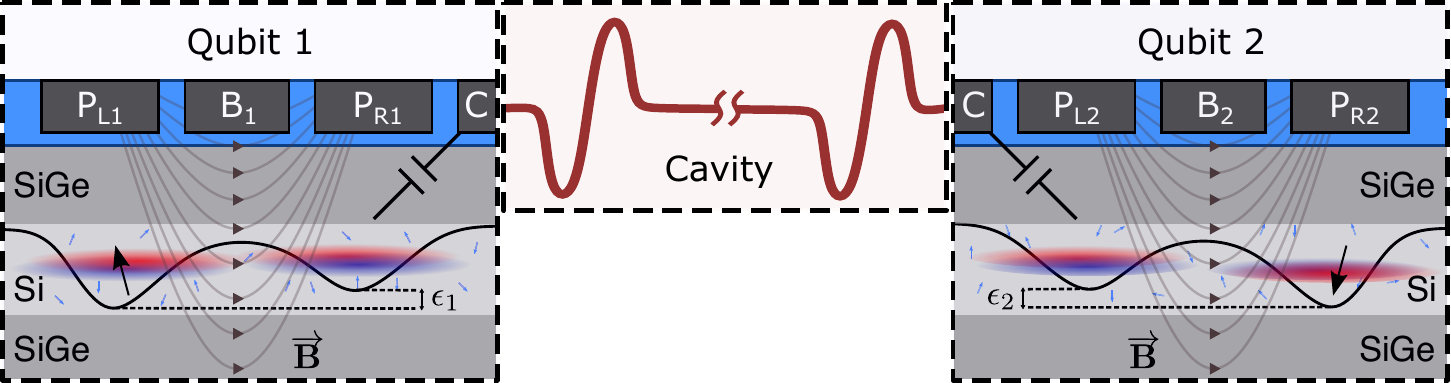}
	\caption{Illustration of a pair of spin qubits coupled via a cavity. Qubits 1 and 2 are each encoded in the spin of a single electron, occupying the space spanned by the left and right orbitals, each having a ground (blue) and excited (red) eigenvalley that may have different interdot tunnel couplings. Micromagnets provide a magnetic field gradient and enable single-qubit control via EDSR. The barrier gate B modulates the interdot tunnel coupling, $t_{c}$, while the plunger gates $P_{L},P_{R}$ control the detuning $\epsilon$. The cavity is capacitively coupled to each qubit via a nearby gate electrode $C$ \cite{Borjans2020}. Impurity $^{29}\mathrm{Si}$ nuclei are depicted with small blue arrows.}
	\label{fig:DeviceCartoonl}
\end{figure*}

Second, we include the presence of valleys, associated couplings, and noise processes \cite{Yang2012}. If the energy splitting between valley eigenstates is insufficiently high, these states have the potential to interfere with gating operations \cite{Culcer2010,Borjans2021}. The impact of valleys on charge-cavity coupling was theoretically examined by Burkard \textit{et al.} and measurements of the cavity response were proposed as a means to characterize valley splitting \cite{Burkard2016}. Subsequent experiments demonstrated the impact of valleys in Si on the cavity response \cite{Mi2017,Mi2018b,Borjans2021}. With their inclusion in the model, we are able to assess the consequences of valleys as a function of the valley splitting.

Finally, in addition to error mechanisms such as relaxation due to electron-phonon coupling, cavity loss, and low- and high-frequency charge noise, we also include low-frequency magnetic noise due to the hyperfine interaction with nuclear spins in the host lattice \cite{Burkard2021}. Hyperfine coupling was not considered in past theoretical work. As we will demonstrate, these factors have significant -- and competing -- effects on gate performance, resulting in tradeoffs that must be navigated to identify the optimal device characteristics and operating parameters. Based on our analysis, we highlight the sources of performance degradation that are most critical to address in future devices. Ultimately, improvements to the iSWAP gate fidelity will most readily be obtained by increasing the spin-photon coupling rate $g_s$ and reducing the cavity decay rate $\kappa$. 

\section{Model}
\label{sec:Model}
An iSWAP gate is defined as 
\begin{flalign*}
	{\rm iSWAP}=\left[\begin{array}{cccc}
		1 & 0 & 0& 0\\
		0 & 0 & i& 0\\
		0 & i & 0& 0\\
		0 & 0 & 0& 1
	\end{array}\right].
\end{flalign*}
This is a unitary operation that may be generated by a two-qubit interaction Hamiltonian of the form $\hat{H}_{\mathrm{int}}=-g (\sigma_{x}\sigma_{x}+\sigma_{y}\sigma_{y})$ evolving a system for a time $t$ such that $\sin(2gt)=1$~\cite{Krantz2019}. As we will show, this interaction can be engineered using the system we consider here,
consisting of two DQDs that are capacitively coupled to a half-wavelength superconducting cavity with resonance frequency $\omega_c$. A schematic of the device is shown in Fig.~\ref{fig:DeviceCartoonl}. The electrostatic potential of each DQD is controlled by the voltages applied to gate electrodes, with the two primary control knobs of interest for each DQD being the interdot energy level detuning $\epsilon$ and the interdot tunnel coupling $t_{c}$. Each DQD contains a single electron that interacts with the electric field of the cavity through the electric dipole interaction. We encode a qubit using the spin degree of freedom of a single electron \cite{Loss1998}. The magnetic field gradient provided by a proximal micromagnet generates the synthetic spin-orbit coupling that facilitates single-qubit operations \cite{Pioro-Ladriere2008} and the spin-cavity coupling required for the two-qubit interaction \cite{Mi2018,Benito2019}.

The full system is modeled by the Hamiltonian
\begin{equation}
\op{H}_{\mathrm{tot}} = \sum_{i=1,2} \op{H}_{\mathrm{DQD}}^{i}+\op{H}^{i}_{\mathrm{DQD-C}} + \op{H}_{\mathrm{C}},
\label{eq:Htot}
\end{equation}
where $\op{H}_{\mathrm{DQD}}^i$ describes DQD $i$, $\op{H}_{\mathrm{C}}$ models the cavity field, and $\op{H}^{i}_{\mathrm{DQD-C}}$ captures the interaction between a DQD and the cavity field. We now describe each term in detail.

A given DQD has a Hamiltonian $\op{H}_{\mathrm{DQD}}^{i}$ describing three degrees of freedom, namely (1) the orbital state $\vert L \rangle, \vert R \rangle$, (2) the spin state $\vert \! \downarrow \rangle, \vert \! \uparrow \rangle$, and (3) the valley state the electron occupies $\vert v_{-}\rangle, \vert v_{+} \rangle$. The orbital states simply describe the spatial localization of the wavefunction in the left or right minimum of the DQD confinement potential. 
For a given quantum dot $i$, the valley eigenstates are linear combinations of states that have support at the pair of conduction band minima of silicon corresponding to the direction of confinement.  The states at these minima are characterized by Bloch wavevectors of opposite sign and are degenerate in bulk silicon; the confining potential breaks translational symmetry and couples these states, resulting in a pair of linear combinations of these states $v_{+}$ and $v_{-}$ that are split by an energy that depends strongly on the details of confinement and disorder in the quantum well \cite{Friesen2007,Culcer2010b,Neyens2018}.   Our convention here is to denote $v_{+}$ ($v_{-}$) as the ground (excited) valley states. The tunnel coupling between quantum dots depends, in general, on the valley character of the states in each dot and their relative phase of the contributing valley states in  $v_{+}$  and $v_{-}$\cite{Borjans2021}. The tunnel coupling $t_{c} = \langle \psi_{L,\pm} \vert H \vert \psi_{R,\pm} \rangle$ between counterpart valley eigenstates may be different in magnitude from that between ground and excited valley eigenstates $t_{c}' = \langle \psi_{L,\pm} \vert H \vert \psi_{R,\mp} \rangle$, depending on the difference in valley character of $v_{+}$  and $v_{-}$ between dots \cite{Borjans2021}.

The DQD Hilbert space in this model is eight-dimensional, with the Hamiltonian
\begin{eqnarray}
\op{H}_{\mathrm{DQD}} & = &\frac{\epsilon}{2}\op{\tau}_z+t_{c} \op{\tau}_x +t_{c}' \op{\tau}_x\op{\nu}_x \nonumber \\
& & + \frac{B_z}{2}\op{\sigma}_z + \frac{b_x}{2}\op{\sigma}_x\op{\tau}_z \nonumber\\
& & + \frac{1}{2}\op{\nu}_z ( v_{L} \vert L \rangle \! \langle L \vert + v_{R} \vert R \rangle \! \langle R \vert) \nonumber \\
& & + \op{\nu}_{x} \op{\sigma}_{x} \left( \frac{\lambda_{sv,L}}{2} \vert L \rangle \! \langle L \vert + \frac{\lambda_{sv,R}}{2} \vert R \rangle \! \langle R \vert \right), \label{eq:HDQD}
\end{eqnarray}
where the Pauli matrices $\op{\tau}_{k}$, $\op{\sigma}_{k}$, and $\op{\nu}_{k}$ act on orbital, spin, and valley sectors, respectively. For simplicity we consider DQD's 1 and 2 to be identical and have omitted the DQD index $i$, but in general all parameters may be distinct for each DQD. The first and second terms of Eq.~\ref{eq:HDQD} are those of a single electron charge qubit \cite{Hayashi2003}. The third term accounts for the fact that the eigenvalley for each dot may have different valley character, resulting in a tunnel coupling $t_{c}'$ between the ground eigenvalley of one dot and the excited eigenvalley of the other dot \cite{Borjans2021}. The fourth and fifth terms describe the uniform applied magnetic field $B_{z}$ as well as the transverse magnetic field gradient $b_{x}$ due to the micromagnet. The sixth term describes the valley splitting $v_{L}, v_{R}$ for each dot.  We assume identical valley splittings $v_{L}=v_{R}=v$ for each dot in the DQD. In practice, significant variations in the valley splitting are observed \cite{Borselli2011,Borjans2019,Hollmann2020,Chen2021}, though in some cases similar splittings for a DQD may be attained through tuning \cite{Mi2017}. The seventh term describes spin-valley coupling arising from intrinsic spin-orbit coupling \cite{Yang2013}.  Spin-valley hybridization due to this coupling leads to $T_{1}$ ``hot spots'' when valley and spin splitting energies are close to one another \cite{Yang2013,Hao2014,Zhang2020,Jock2022}. Here, we assume identical spin-valley coupling in each dot, $\lambda_{\mathrm{sv},L}=\lambda_{\mathrm{sv},R}=\lambda_{\mathrm{sv}}$. Note that the ratio between the inter- and intra-valley tunnel coupling $t_{c}'/t_{c}$ is understood to depend on disorder intrinsic to each dot as well as its tuning \cite{Borjans2021}, so we do not assume independent control of $t_{c}'$ but rather fix $t_{c}'/t_{c}=0.5$.

\begin{figure*}[t]
	\centering
	\includegraphics[width=2\columnwidth]{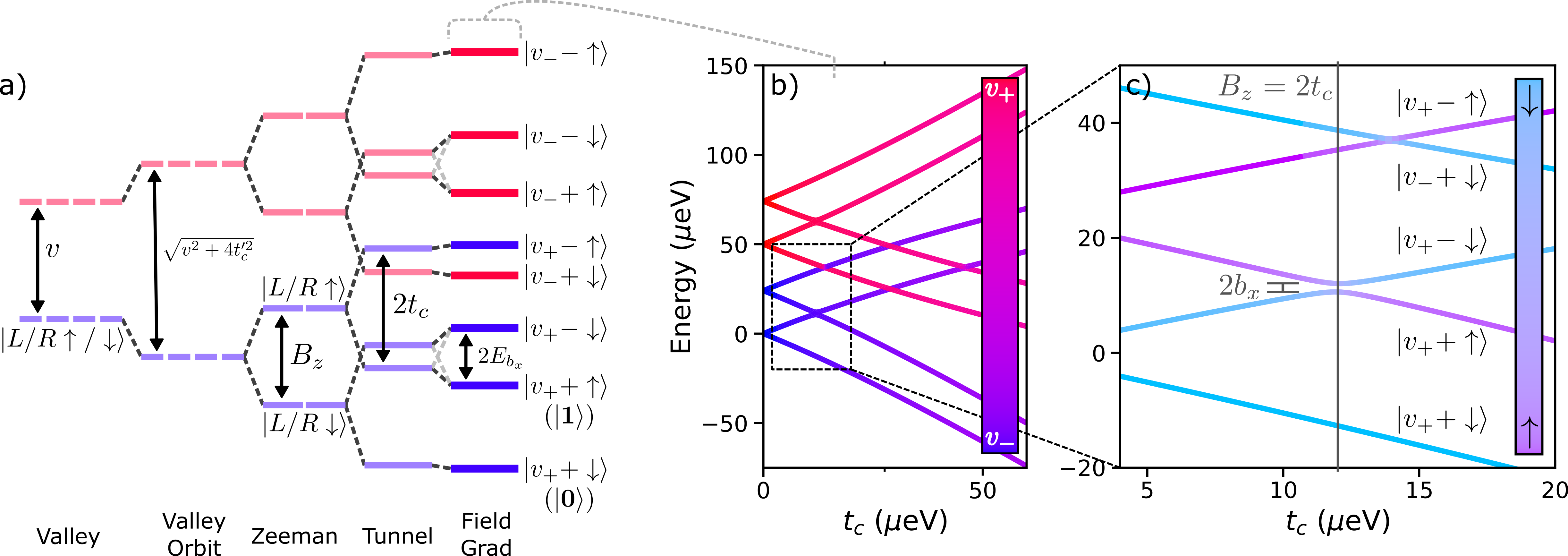}
	\caption{\label{fig:single_DQD} 
	a) Energy level diagram showing the impact of Hamiltonian terms on the states of  the DQD at $t_c\approx 15\mu$~eV. Energy splittings are not to scale and are exaggerated for visibility, and $E_{b_x}=\sqrt{(B_z^2-2t_c^2)+b_x^2}$. b) The DQD energy levels as a function of $t_c$ for fixed $\epsilon=0$ 
	and valley splitting $v=50 \ \mathrm{\mu eV}$. The coloring indicates the valley character of each energy level.  All other parameters are set according to Table \ref{tab:Params}.  c) Magnified region of the center plot with the energy bands colored according to the associated state's spin character.  
	For $t_{c} > B_{z}/2$, the lowest two states constitute the qubit logical basis $\vert v_{+} + \downarrow \rangle = \vert \mathbf{0} \rangle$, $\vert v_{+} + \uparrow \rangle = \vert \mathbf{1} \rangle$. }
\end{figure*}

Finally, the cavity and DQD-cavity coupling Hamiltonians of Eq. \ref{eq:Htot} are given by \cite{Beaudoin2016}
\begin{eqnarray}
	\op{H}_{\mathrm{C}} & = &\omega_c \op{a}\dg \op{a}, \label{eq:HC} \\
	\op{H}^{i}_{\mathrm{DQD-C}} & = & g_{c}^{i}\op{\tau}_z^{i}(\op{a}+\op{a}\dg), \label{eq:HDQDC}
\end{eqnarray}
where $\op{a}$($\op{a}\dg$) is the bosonic annihilation(creation) operator for the cavity mode, $\omega_c$ is the frequency of the cavity mode, and $g_c^{i}$ quantifies the electric dipole coupling between DQD $i$ and the cavity \cite{Benito2019}. For simplicity we assume the dipole coupling $g_{c}$ is the same for both DQDs, i.e. $g_{c}^{1}=g_{c}^{2}=g_{c}$. In contrast with the investigations of \cite{Benito2017,Benito2019,Warren2019}, we will not assume a dispersive limit of virtual cavity population but rather include the field mode explicitly in all simulations. We truncate the maximum cavity photon number to $n_c \leq 2$, resulting in $\op{H}_{\mathrm{tot}}$ having dimension $8\times 8\times 3=192$ due to the system being the tensor product of two eight-level DQD subsystems and one three-level cavity. Unless otherwise specified, in this work we adopt the units convention that $\hbar = 1$.

\subsection{DQD system}

The Hamiltonian in Eq.~\eqref{eq:HDQD} results in a spectrum of eigenstates with mixed spin, orbit, and valley character.  The energy level diagram in Fig.~\ref{fig:single_DQD}(a) shows the impact of the various terms on the DQD eigensystem. First, valley splitting $v$ separates the lower ($\ket{v_+}$) and upper ($\ket{v_-}$) valley states and the intervalley tunnel coupling $t_c'$ leads to hybridized states of mixed valley character.  When $v$ is large compared to $t_c'$ -- as depicted in the diagram -- the hybridization is weak and the DQD states can be separated into two manifolds with predominantly one type of valley character, shown in blue and red.  
Second, the uniform magnetic field $B_z$ splits the up ($\ket{\uparrow}$) and down ($\ket{\downarrow}$) spin states; we adopt the convention that down spins are lower in energy ($g \approx$ 2 for electrons in Si).  
Third, the tunnel coupling $t_c$ hybridizes the orbital states ($\ket{L}$, $\ket{R}$) of the component dots, resulting in lower energy bonding ($\ket{+}$) and higher energy anti-bonding ($\ket{-}$) states for small values of the detuning $\vert \epsilon \vert \ll t_{c}$. In this work we assume $\epsilon=0$, an operational ``sweet spot'' where the system is first-order insensitive to detuning noise and the two hybridized states have equal $L/R$ character \cite{Vion2002}.  Figure~\ref{fig:single_DQD}(b) shows the energy spectrum as a function of $t_c$, with the energy bands again colored according to valley character. At $t_c=0$, there are four degenerate pairs of orbital states sharing the the same spin/valley character, for each valley the two spins are split in energy from each other by $B_z$, while the two valleys are split from each other by $v$ (here set to 50 $\mu$eV). As the tunnel coupling becomes finite, each pair of orbital states split into $\ket{\pm}$ states; at the point $B_z=2t_c$ the $\ket{+\uparrow}$ and $\ket{-\downarrow}$ states within each valley sector cross in energy, and as $t_c$ (and consequently $t_c'$) becomes significant, the valleys begin to hybridize significantly.

Finally, the magnetic field gradient $b_x$ hybridizes the $\ket{+\uparrow}$ and $\ket{-\downarrow}$ states within each valley sector, the strength of which depends on the mixing angle $\phi=\tan^{-1}\left(\frac{b_x}{\abs{2t_c-B_z}}\right)$\cite{Benito2019}.  Figure~\ref{fig:single_DQD}(c) shows a magnified view of the avoided crossing  of the $\ket{v_++\uparrow}$ and $\ket{v_+-\downarrow}$ states with the energy bands colored according to spin character.  These states now have mixed spin-orbit character, which is crucial to operation of the device; however, due to the small magnitude of $b_x$ these states will be predominantly one spin unless $\phi\sim\pi/2$ very near $B_z=2t_c$.  As such, since we are working in the regime of small $\phi$ with $2t_c>B_z$, we will continue to use the $\uparrow/\downarrow$ labeling for these states.  The two lowest lying states are now $\ket{v_++\downarrow}$ and $\ket{v_++\uparrow}$. We therefore define $\ket{v_++\downarrow}$ =  $\ket{\mathbf{0}}$ and $\ket{v_++\uparrow}$ = $\ket{\mathbf{1}}$ as the two states of the DQD qubit.

\subsection{Cavity-mediated spin-spin interactions}
\begin{figure}
	\centering
	\includegraphics[width=0.99\columnwidth]{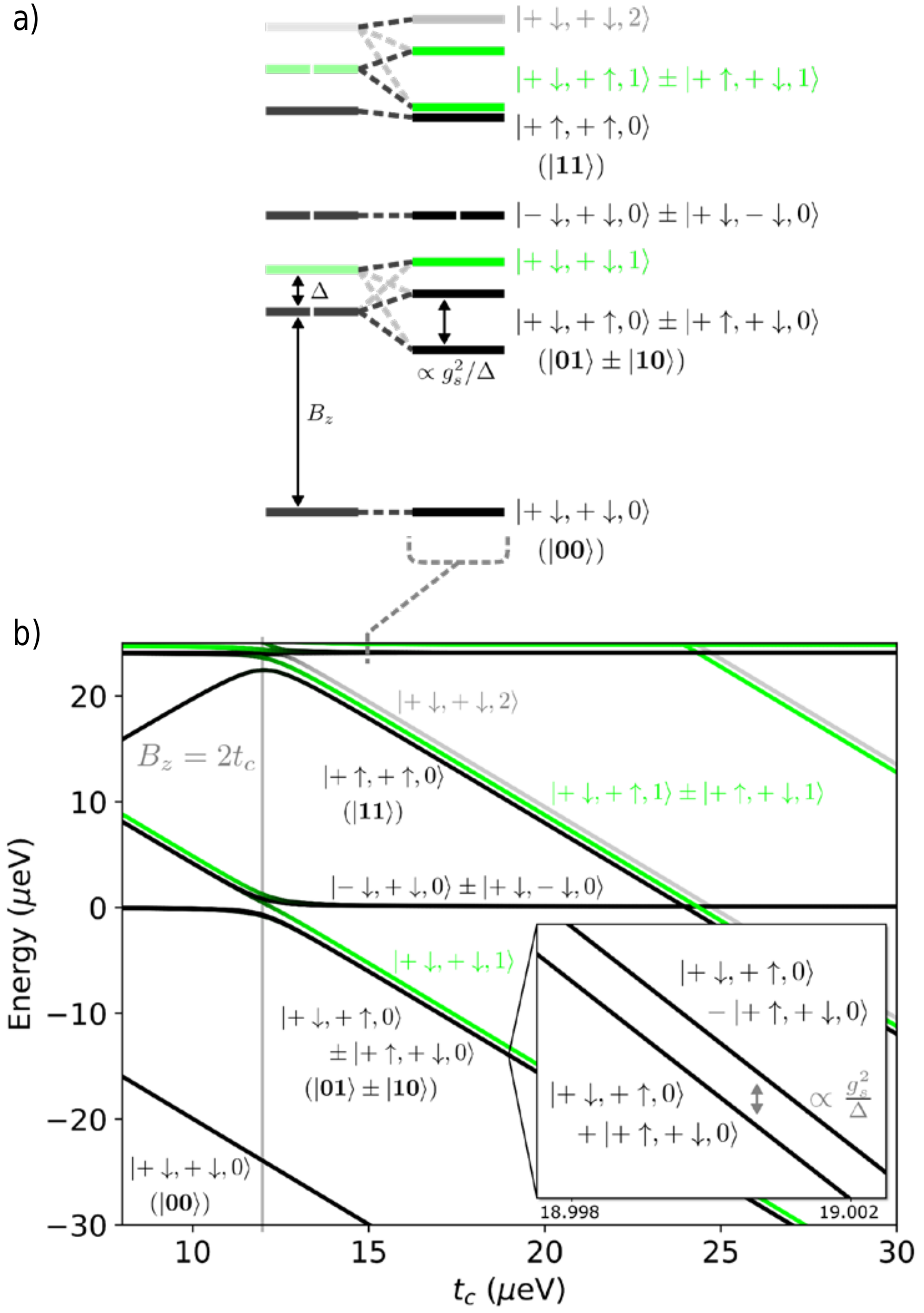}
	\caption{\label{fig:e_gate}(a) Energy levels for the combined spin-photon system with both DQDs. (b) Energy levels as a function of $t_{c}^{1} = t_{c}^{2} = t_c$ with $\omega_c$ varied such that $\Delta=200$ MHz, and $v\rightarrow\infty$.  As in Fig. \ref{fig:single_DQD}, the logical qubit states are $\vert + \downarrow \rangle = \vert \mathbf{0} \rangle$, $\vert + \uparrow \rangle = \vert \mathbf{1} \rangle$. Black indicates spin character (zero cavity occupation) while green indicates $n_c$ = 1 and gray indicates $n_c$ = 2.  The labels denote the primary character of the two DQDs. We emphasize that as the $t_c=B_z/2$ point (vertical gray line) is approached, the states acquire mixed character as shown in Fig. \ref{fig:single_DQD}. }
\end{figure}

These qubit states can now be made to interact with the cavity. The essential idea is that, since the electron spin couples to the orbital degree of freedom through the engineered magnetic field gradient $b_x$ and the orbital degree of freedom interacts with the cavity through electric dipole coupling $g_c$, the cavity field couples to spin-spin transitions between the $\ket{\mathbf{0}}$ and $\ket{\mathbf{1}}$ states of each DQD. The states of the complete system can be constructed from tensor products of  eigenstates of the two DQDs and the cavity $\ket{v_\pm \pm \updownarrow}_{1} \ket{v_\pm \pm \updownarrow}_2 \ket{n}_c \rightarrow \ket{v_\pm \pm \updownarrow, v_\pm \pm \updownarrow,n}$. In what follows, we are concerned primarily with states in the lowest energy valley sector and will omit the valley indices of the states for brevity.  In particular, low-lying energy states of the system comprising qubit states of the two DQDs now form a two-qubit system; we will label the two-qubit states $\ket{jk}$ with $j,k\in \lbrace \mathbf{0},\mathbf{1} \rbrace$, so that $\ket{\mathbf{00}}=\ket{+\downarrow,+\downarrow,0}$,$\ket{\mathbf{01}}=\ket{+\downarrow,+\uparrow,0}$, $\ket{\mathbf{10}}=\ket{+\uparrow,+\downarrow,0}$, and $\ket{\mathbf{11}}=\ket{+\uparrow,+\uparrow,0}$.  The energy level diagram of the combined system is given in Fig. \ref{fig:e_gate}(a),  showing the impact of DQD-cavity coupling $g_c$ on the overall eigenstates of the system. We highlight two sets of states specifically. First, we note that immediately above the global ground state of the system are the $\ket{\mathbf{01}}/\ket{\mathbf{10}}$ and the $\ket{+\downarrow,+\downarrow,1}$ states, separated in energy by $\Delta=\omega_c-(E_1-E_0)$, where $\omega_{c}$ is the cavity photon energy and $E_1-E_0$ is the DQD transition energy. When $\Delta$ is small, the coupling $g_c$ hybridizes the cavity and DQD states, the latter of which form bonding and anti-bonding states split in energy by 2$g_s\sim 2g_c\sin\left(\phi/2\right)$. Second, we note the three states at around $2\omega_{c}$: $\ket{\mathbf{11}}$, $\ket{+\uparrow,+\downarrow,1}/\ket{+\downarrow,+\uparrow,1}$, and $\ket{+\downarrow,+\downarrow,2}$. As in the prior case the $\ket{+\uparrow,+\downarrow,1}/\ket{+\downarrow,+\uparrow,1}$ and $\ket{+\downarrow,+\downarrow,2}$ states hybridize, with the former pair forming split bonding and anti-bonding states. These further hybridize with the $\ket{\mathbf{11}}$ state, pushing it lower in energy. 

Figure \ref{fig:single_DQD}(b) shows the dependence of these states' energy and character on $t_c$.  We have assumed in this plot that $\omega_{c}$ is modulated to achieve constant $\Delta$.  At $B_z=2t_c$ the avoided crossings in the underlying DQD states are  visible.  In addition, strong spin-orbit mixing results in greater effective  $g_s$ and hybridization of the DQD states with the cavity, complicating the structure of the avoided crossings for the overall system.  The magnified inset highlights the splitting between the states, which is proportional to $g_s^2/\Delta$ and sets the time scale for the evolution of a state in $\ket{\mathbf{01}}$ to $i\ket{\mathbf{10}}$ and $\ket{\mathbf{10}}$ to $-i\ket{\mathbf{01}}$.  Conversely, the states $\ket{\mathbf{00}}$ and $\ket{\mathbf{11}}$ correspond to eigenstates of the system and will be stable over such an interval.  Thus we see that for a given $t_c$ the system is capable of performing an iSWAP operation on the two qubit subsystem over an appropriate duration.  This indicates device operation consisting of an ``off'' state with tunnel coupling $t_c^{\rm MAX}$ set very large, so that $\phi\approx 0$ and the DQDs and cavity are essentially uncoupled.  The iSWAP is performed by ramping $t_c$ to a target value $t_c^0$ where $\phi$ is finite, allowing the system to evolve through the iSWAP operation, and ramping $t_c$ back to the uncoupled regime.  An important caveat to this is that changes to $t_c$ must be sufficiently slow; as is evident from Fig. \ref{fig:e_gate}(b), especially near the avoided crossing region at $B_z=2t_c$, the $t_c$ dependent hybridization of the DQD states with the cavity and small energy splittings can result in Landau-Zener transitions to excited cavity states outside the two-qubit subspace~\cite{Warren2019}. Below, we will explore in detail how the severity of this effect depends on the $t_c$ control schedule.

In our following analyses, we choose the Hamiltonian parameters indicated in Table \ref{tab:Params}. To probe the relative influence of valley splitting, DQD-cavity dipole coupling, and cavity loss rate on two-qubit errors, we allow for these parameters to vary across experimentally realistic ranges of values.
\begin{table}[h]
	\centering
	\begin{tabular}{|c|c|c|}
		\hline \textbf{Description} & \textbf{Variable} & \textbf{Value} \\
		\hline
		\hline Longitudinal B-field & $B_{z}$ & $24 \ \mathrm{\mu eV} (\sim 200)$mT \\
		\hline Transverse B-field gradient & $b_{x}$ & $1.5 \ \mathrm{\mu eV} (\sim 1.3)$mT \\
		\hline Interdot detuning & $\epsilon$ & $0$ \\
		\hline Operating tunnel coupling & $t_c^0$ & variable \\
		\hline Maximum (inactive state) & $t_c^{\rm MAX}$  & variable \\
		tunnel coupling & & \\
		\hline Valley splitting & $v$ & variable \\
		\hline Inter-/intra-valley tunnel & $t_{c}'/t_{c}$ & $1/2$ \\
		coupling ratio &  &  \\
		\hline Spin-valley coupling & $\lambda_{sv}$ & $1 \ \mathrm{\mu eV}$ \\
		\hline DQD-cavity electric dipole & $g_{c}$ & variable \\
		\hline DQD-cavity detuning &$\Delta$& variable\\
		\hline Cavity frequency  & $\omega_c$ & $E_e(t_c^0)-E_g(t_c^0) +\Delta$ \\
		\hline Cavity loss rate & $\kappa$ & variable \\
		\hline Phonon relaxation rate & $\gamma_{\rm ph}$ & variable \\
		\hline Charge noise dephasing rate & $\gamma_{\rm ch}$ & variable \\
		\hline Intervalley phonon & $\gamma_{\rm vph}$ & $0.5\gamma_{\rm ph} $ \\
		relaxation rate & & \\
		\hline Valley dependent charge & $\gamma_{\rm vch}$ & $0.01\gamma_{\rm ch} $\\ 
		noise dephasing rate &  &  \\
		\hline Low-frequency variation in & $\delta_{B_z}$ & $100 \ $kHz \\ longitudinal magnetic field &  &  \\
		\hline Low-frequency variation in & $\delta_{\epsilon}$ & $0.2 \ \mathrm{\mu eV}$\\
		interdot detuning &  & \\
		\hline
	\end{tabular}
	\caption{Hamiltonian and decoherence parameters considered in our analysis.  $E_e(t_c^0)-E_g(t_c^0)$ is the energy difference between the ground ($\ket{v_++\downarrow}$ =  $\ket{\mathbf{0}}$) and first excited states ($\ket{v_++\uparrow}$ = $\ket{\mathbf{1}}$) composing a qubit of a DQD. }
	\label{tab:Params}
\end{table}

\subsection{Decoherence mechanisms}
The three sources of decoherence that we consider here are (1) charge noise, (2) phonon relaxation, and (3) magnetic noise due to the contact hyperfine interaction with nuclear spins in the host lattice. Quasi-static charge noise on detuning and interdot tunnel coupling was considered in Ref.~\cite{Warren2019}, and in Refs.~\cite{Benito2017,Benito2019} dephasing and relaxation due to charge noise and phonons were considered in the context of spin-cavity coupling.

In this work, we include both high-frequency and low-frequency (quasi-static) noise. We include the effect of high-frequency noise by propagating the dynamics according to a master equation \cite{Breuer2002}
\begin{flalign}
    \dot{\rho} = -i[\op{H}(t),\rho] + \sum_{m,n,x} \op{L}_{x}^{mn} \rho \op{L}_{x}^{mn \dagger} - \frac{1}{2} \lbrace \op{L}_{x}^{mn \dagger} \op{L}_{x}^{mn}, \rho \rbrace
\end{flalign}
that includes Lindblad terms of the form 
\begin{flalign}
    \op{L}_{x}^{mn}(t_c)=\vert E_{m} \rangle \! \langle E_{m} \vert \op{H}_{x} \vert E_{n} \rangle \! \langle E_{n} \vert \beta(\omega_{mn}(t_c))\label{eq:hf_noise},
\end{flalign}
where $\lbrace \vert E_{m} \rangle \rbrace$ are instantaneous eigenstates at a given $t_c$ and $\omega_{mn} = E_{m} - E_{n}$ are corresponding energy differences. The function $\beta(\omega)$ characterizes the noise spectrum of the bath, and the operator $\op{H}_{x}$ corresponds to the system part of each system-bath interaction operator. For the present work, we use simplified spectral functions that capture the qualitative behavior of the type of bath.  

In the case of charge noise, we assume the phenomenological model inspired by the spin-boson model \cite{Leggett1987}
\begin{flalign}
	\op{H}_{\rm ch}=\gamma_{\rm ch}\left(\op{\tau}_z+\op{\tau}_x\right)+\gamma_{\rm vch}\op{\upsilon}_z,
\end{flalign}
with $\gamma_{\rm ch}$ the strength of noise in both the DQD  energy level  detuning $\epsilon$ and the tunnel coupling $t_c$, and $\gamma_{\rm vch}$ being the noise strength in valley splitting $v$. Since the power spectrum of charge noise is expected to drop off significantly with frequency and most of the transition frequencies of interest are roughly similar, the associated $\beta(\omega)$ is chosen to be a unit normal delta function, so that only dephasing terms remain in any basis.

We model the low-frequency noise component, treated as quasi-static, by sampling from a Gaussian distribution of the detuning $\epsilon$ on each DQD with standard deviation $\delta_{\epsilon}$ and assuming that the detuning to be constant over the timescale of a given circuit realization.

We expect low-frequency charge noise to also act on the tunnel coupling, \cite{Warren2019}, and in principle this may be incorporated in a similar fashion as noise on $\epsilon$ so that $t_c^0$ is sampled from a distribution with standard deviation $\delta t_c$.  However, we do not consider this noise mechanism in detail here, as we expect it to be negligible according to the following argument. 
Considering a typical lever arm for detuning on the order of 100 $\mathrm{\mu eV/mV}$ \cite{Zajac2016}, the quasi-static gate-referred noise strength comparable to the detuning noise level $\delta_{\epsilon} = 0.2 \ \mathrm{\mu eV}$ of Table \ref{tab:Params} would be approximately 2 $\mathrm{\mu V}$. In Ref \cite{Borjans2021}, a variation in tunnel coupling on the order of 1 $\mathrm{\mu eV/mV}$ was observed as a function of voltage on the barrier gate electrode. Assuming the same gate-referred noise level on the barrier gate as for the detuning gates, this would correspond to $\delta t_c$ of the order of a few $\mathrm{neV}$. We find that this magnitude of fluctuation in $t_{c}$ should result in a small enough iSWAP error that we may ignore it in the present case.  However, a definitive assessment requires a more thorough analysis incorporating details of device electrostatics. 

We assume that relaxation in the DQDs is due to electron-phonon coupling \cite{RidleyBook}.
The phonons are assumed to directly couple only to the dot dipole and intra-dot valley dipoles; the phenomenological Hamiltonian we use in Eq.~\eqref{eq:hf_noise} is
\begin{flalign}
	\op{H}_{\mathrm{ph}}=\gamma_{\rm ph}\op{\tau}_z +\gamma_{\rm vph}\op{\upsilon}_z,
\end{flalign}
where $\gamma_{\rm ph}$ is the intravalley phonon-mediated relaxation rate and $\gamma_{\rm vph}$ is the intervalley phonon-mediated relaxation rate. While a realistic $\beta(\omega)$ is expected to grow monotonically from $\omega=0$ \cite{RidleyBook}, we are mostly concerned with processes with similar energy and will assume a simplified form. We will take $\beta(\omega)=1$ for $\omega>0$ and $\beta(\omega)=0$ for $\omega\le0$, as we are assuming only relaxation processes due to low temperature. Additionally, we include magnetic noise due to hyperfine interactions with spinful nuclei in the host lattice. We treat this noise in the low frequency limit, making the quasi-static approximation by sampling from Gaussian distributions of $B_z$ on each DQD with standard deviation $\delta_{B_z}$.

\begin{figure*}[t]
	\centering
	\includegraphics[width=2\columnwidth]{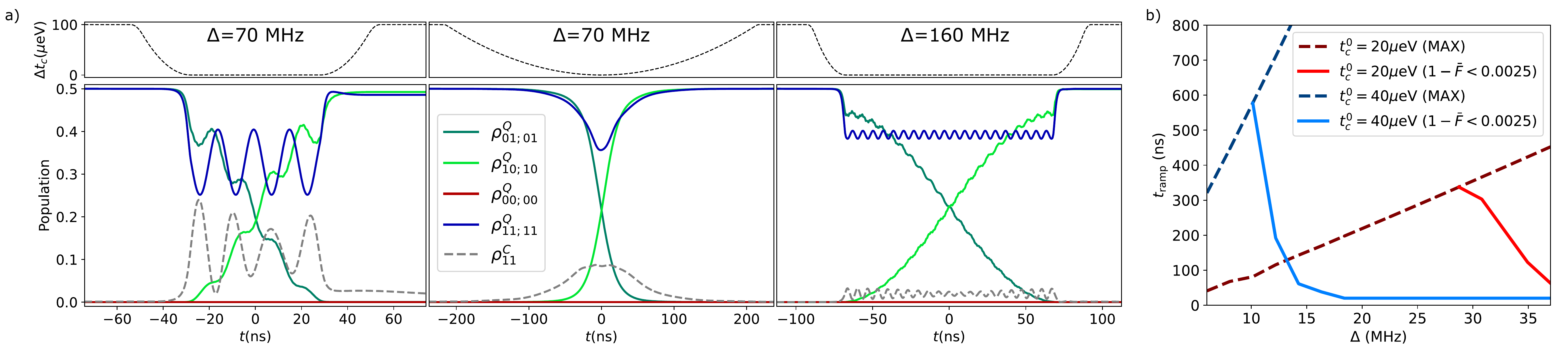}\label{fig:ud_all}	
	\caption{(a) Population dynamics of a decoherence-free iSWAP gate for different qubit-cavity detunings $\Delta$ and  $t_c$ pulse profiles  with $g_c$ = 50 MHz, $t_c^0$ = 13 $\mu$eV and $t_c^{\rm MAX}=t_c^0+100$ $\mu$eV. $\rho^Q$ is the density matrix in the qubit basis (with cavity degrees of freedom traced over) and the indices are the qubit states as described above; $\rho_C$ is the density matrix of the cavity with DQD degrees of freedom traced over. These plots assume an initial state $\frac{\ket{\mathbf{11}}+\ket{\mathbf{01}}}{\sqrt{2}}$. 
	The first two panels show results for $\Delta$ = 70. The left panel shows the system evolution with a 20 ns ramp duration at the beginning, yielding infidelity $1-\bar{F}$  = 0.019, and the center shows the evolution with a ramp duration of $\sim$ 212 ns giving $1-\bar{F}=0.0056$. The right panel shows results for $\Delta$ = 160 MHz with a ramp time of 20 ns yielding $1-\bar{F}=0.0032$. (b) Ramp time required to achieve $1-\bar{F}=0.0025$ (if possible) as a function of $\Delta$ for $g_c$ = 100 MHz at $t_c^0$ = 20 $\mu$eV and $t_c^0$ = 40 $\mu$eV.  The dashed MAX lines represent the maximum ramping duration possible given the required gate operating time; i.e.~when the ascending ramp immediately follows the descending ramp. Where the solid lines are absent, the minimum infidelity exceeds the 0.0025 threshold. The minimum ramp time in all cases was arbitrarily set to 20 ns. \label{fig:unitary_dyn}}
\end{figure*}

\section{Decoherence-free iSWAP}
We first consider a scenario where no decoherence is present and the evolution of the entire system  is purely unitary.  
In Fig.~\ref{fig:unitary_dyn} we plot the population dynamics of the iSWAP gate on an arbitrary initial state for abruptly and smoothly modulated $t_c$ for two different system-cavity detunings $\Delta$.

We show the populations of the DQDs (traced over the cavity modes) in terms of the DQD components of the qubit basis, with, \eg $\rho_{01:01}^{Q} = \langle + \! \downarrow, \! + \! \uparrow \! \vert \mathrm{Tr_{C}}\left[\rho\right] \vert + \! \downarrow, \! + \! \uparrow \rangle$, as well as the population in the first excited state of the cavity (traced over DQD degrees of freedom) $\rho^C_{11}$.  The cases shown include one where the cavity is near resonance with the qubit transitions, so that the qubit states are strongly hybridized with the cavity, and one in the dispersive regime where the cavity is strongly detuned from the qubit transitions and cavity states are minimally occupied.     For a sudden  modulation of $t_{c}$ and small detuning, as shown in Fig. \ref{fig:unitary_dyn}(a) (left panel), the population of the cavity is significant and the abrupt $t_c$ pulse results in large oscillations in the qubit state populations due to Landau-Zener transitions to states with higher cavity occupations not adiabatically connected to qubit states. Note that the oscillations in $\rho^Q_{01:01}$ and $\rho^Q_{10:10}$ are not at the same frequency as $\rho^Q_{11:11}$ owing to the slightly different impact of the cavity coupling on these states. Since the oscillations are not quite at the same frequency, it is not possible for the gate operation to be timed to ensure high fidelity operations; while the return ramp can result in transitions back to the correct eigenstates, this process is not perfect.  As expected, a slower pulse as in Fig. \ref{fig:unitary_dyn}(a) (center panel) prevents Landau-Zener transitions to higher lying states, and the fidelity is much improved.  At higher $\Delta$, as we approach the dispersive limit as shown in Fig. \ref{fig:unitary_dyn}(a) (right panel), the cavity population is minimal even with a rapid ramping of $t_c$, though clear oscillations are still visible corresponding to minimal Landau-Zener behavior.  The impact on fidelity is significantly reduced as it is possible to more closely match the oscillation period with the gate duration.  
To see the impact of ramp duration more clearly, in Fig.~\ref{fig:unitary_dyn}(b) we plot the ramp duration needed to achieve a given  infidelity for different values of $t_c^0$.  For small $\Delta$ in the near-resonant regime it is not possible to ramp slowly enough to avoid significant Landau-Zener transitions.  As the dispersive regime is entered and some threshold $\Delta$ is reached, the required ramp duration falls off rapidly; in this regime, the non-qubit states are sufficiently separated from the qubit states such that Landau-Zener transitions can be easily avoided even for fast ramps.

As is evident in these simulations, it is essential to ensure an appropriately smooth pulse profile in order to maximize the fidelity of the gate, especially in the near-resonant and intermediate limits. Thus, in computing the performance of iSWAP configurations in the following analyses we optimize the control pulse profile in addition to the overall gate duration. 
	
\section{Noisy iSWAP}
\begin{figure}
	\centering
	\includegraphics[width=1\columnwidth]{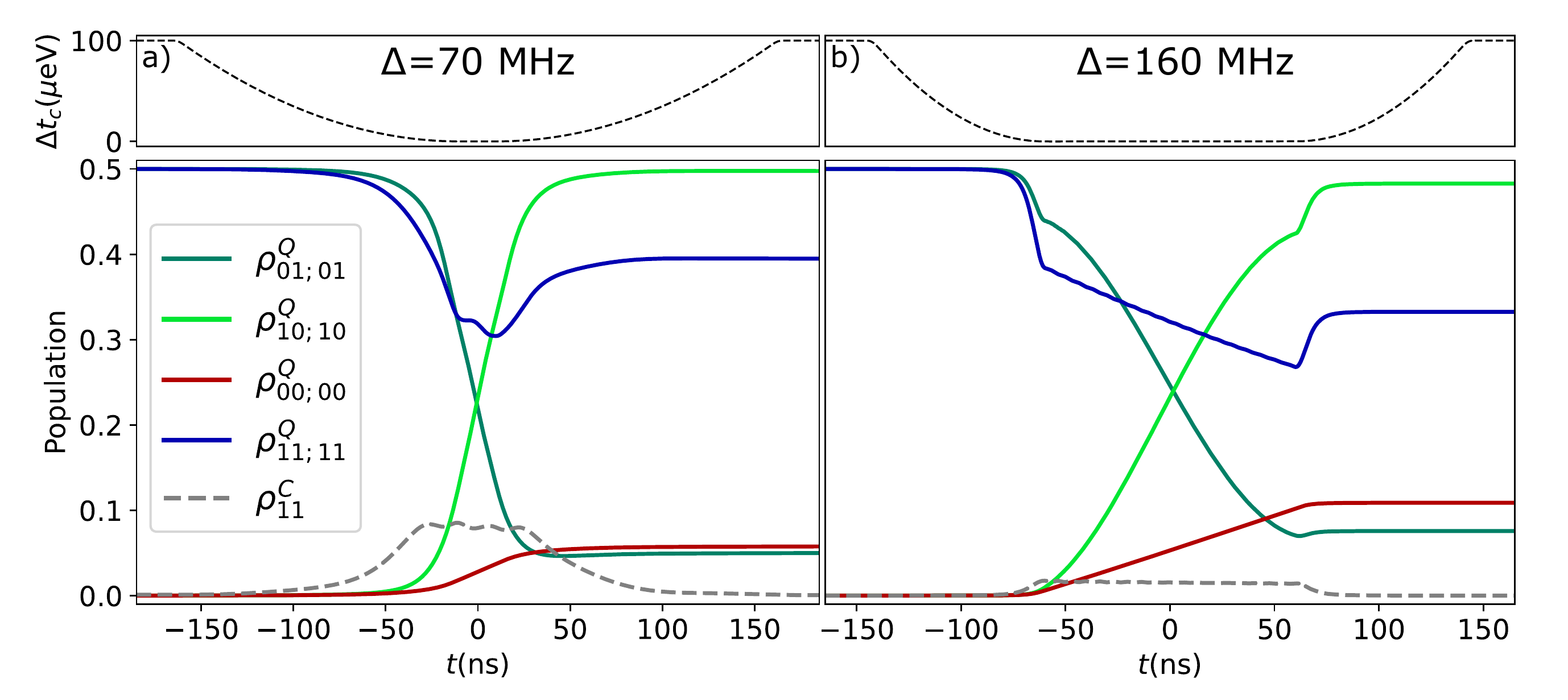}\label{fig:nud_all}
	\caption{Population dynamics of noisy (high-frequency only) iSWAP gates for the different system-cavity detunings in Fig.~\ref{fig:unitary_dyn}.  Pulse profiles for $t_c$ are chosen to minimize Landau-Zener transitions during gate operation.  In all cases $g_c$ = 50 MHz, $\gamma_{\rm ph}=\gamma_{\rm ch}$ = 8 MHz, $\kappa$ = 1 MHz, and the initial state of the joint qubits + cavity system is $\frac{\ket{\mathbf{11}}+\ket{\mathbf{01}}}{\sqrt{2}} \otimes \ket{0}$.  On the left, the infidelity is 0.116; on the right it is 0.195, due to the increased time over which decoherence is able to impact the system. \label{fig:nonunitary_dyn}}
	\end{figure}
We now turn our attention to simulations that include decoherence processes.  We will continue to assume a very large ($\sim 1 \ \mathrm{meV}$) valley splitting  for the moment. In Fig.~\ref{fig:nonunitary_dyn} we investigate the impact of decoherence due to high-frequency processes. We note that for the given parameters, despite the greater cavity population and cavity loss during the gate operation in the near-resonant regime, the fidelity is highest here due to a shorter operating time that limits exposure to phonon and charge noise.  This highlights the significance of the characterization of the control schedule dependence and behavior of the system in the near-resonant regime in the preceding section: while a smooth, slow ramp of $t_c$ reduces loss to the Landau-Zener behavior, it is nonetheless important to ramp as quickly as possible to minimize decoherence.

\begin{figure}
		\centering
		\includegraphics[width=\columnwidth]{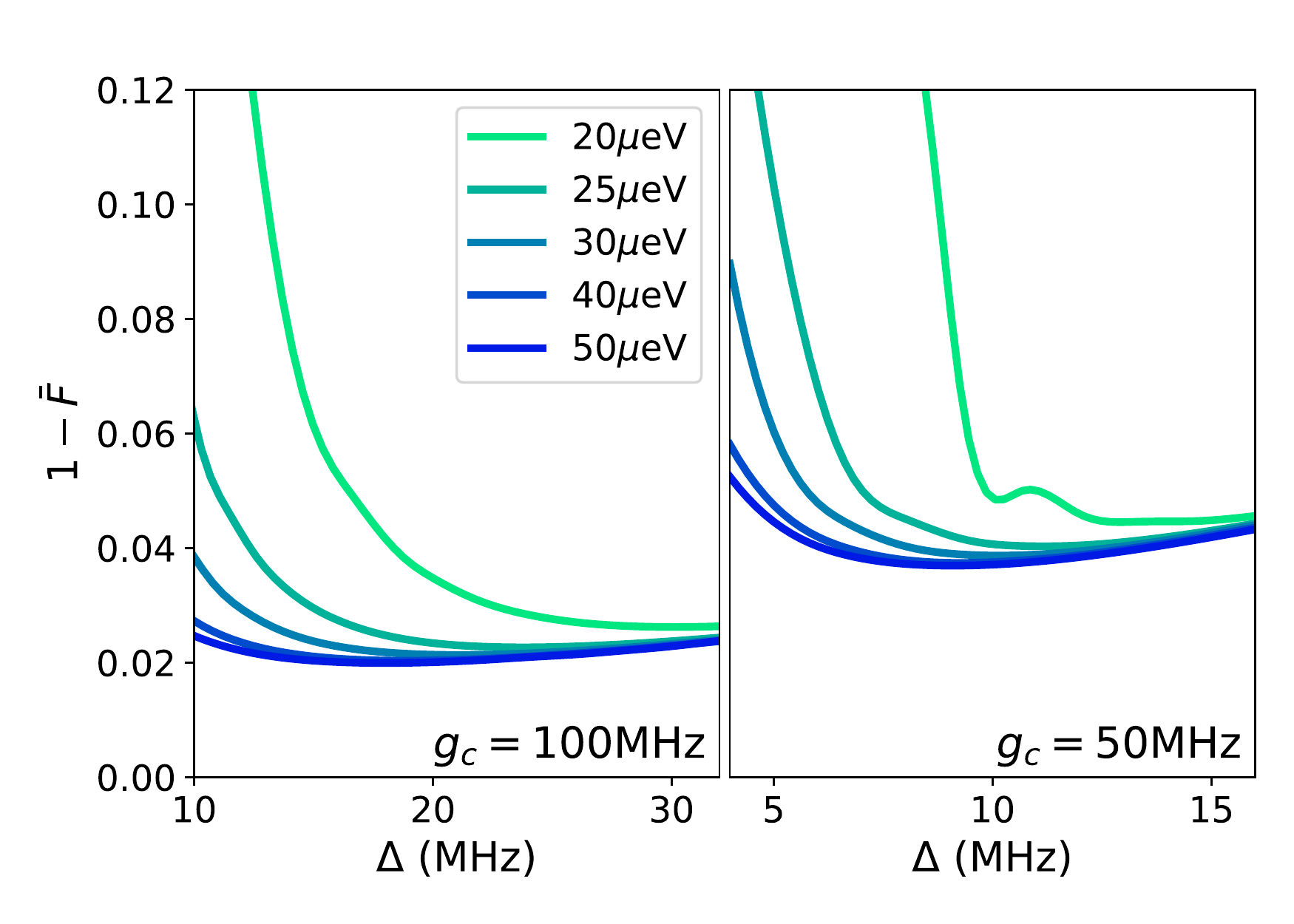}
		\caption{ Average infidelity for a cavity-mediated $\mathrm{iSWAP}$ gate in the presence of only high-frequency noise for the single valley case for $g_c$ = 100 MHz (left) and $g_c$ = 50 MHz (right), with $\gamma_{\rm ph}=\gamma_{\rm ch}$ = 4 MHz, and $\kappa$ = 0.1 MHz. Each curve corresponds to a different choice of $t_c^0$. At small $\Delta$, infidelity is high due to non-adiabaticity and for large $\Delta$ infidelity is limited by the decoherence during the longer time intervals required for gate operation. \label{fig:nFvDelta}}
	\end{figure}
In order to understand this tradeoff more fully, we consider the dependence of fidelity on additional parameters.  First, we identify the optimal control sequence for a given $\Delta$, $g_c$, $t_c^{\rm 0, MAX}$, and decoherence strengths.  In Fig.~\ref{fig:nFvDelta} we show the average infidelity as a function of $\Delta$ for different $t_c^0$ for two parameter sets excluding low-frequency noise.  In general, for small $\Delta$ the infidelity is very high, falling quickly to a minimum as $\Delta$ is increased and approaches the dispersive regime, increasing slowly thereafter due to longer operating times and greater exposure to decoherence.   The origin of the infidelity for small $\Delta$ depends on the cavity loss $\kappa$; when cavity loss is high (not shown) it is responsible for the infidelity due to increased cavity population, however, in the case of low $\kappa$, the fidelity is lost when the implied gate operating time is too short to allow for adiabatic ramping of $t_c$. In contrast with previous results where operation in the dispersive limit was assumed~\cite{Benito2019}, we find that infidelity minima are not invariant with respect to the choice of $t_c^0$.  The impact of effects excluded by the dispersive-regime approximation limit the fidelity for high values of spin-charge mixing (i.e. small $t_{c}^0$), converging for high $t_c$ in the limit where the qubits are primarily spin-like.  Additionally, the optimal value of $\Delta$ depends considerably on the choice of $t_{c}^0$.  In the high spin-charge mixing limit, the optimal $\Delta$ falls well short of the dispersive limit as noise in the qubit system exerts a more pronounced effect and favors shorter gate operating times.  In the spin-like limit of higher $t_{c}$, the reduced sensitivity to charge noise allows for longer gate duration and access to the dispersive regime, where cavity loss can also be minimized. 

\begin{figure*}
		\centering
\includegraphics[width=2\columnwidth]{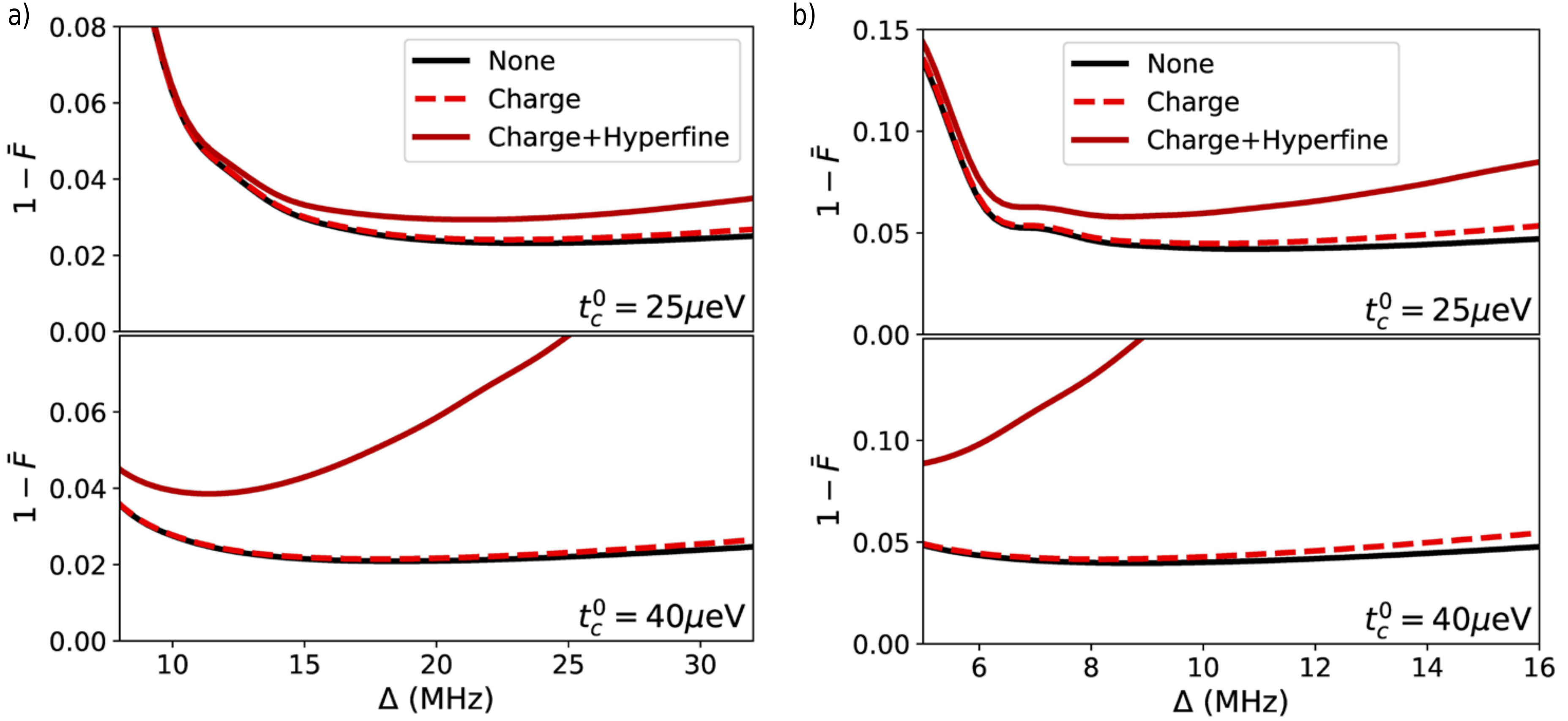}
		\caption{Simulated infidelity with the inclusion of low-frequency noise using the quasi-static approximation for different values of  charge-cavity coupling:  a) $g_c$ = 100 MHz  and b) $g_c$ = 50 MHz. Other parameters are $\gamma_{\rm ph}=\gamma_{\rm ch}$ = 4 MHz, $\kappa$ = 0.1 MHz for both cases. Charge noise has a minimal impact, but hyperfine noise is significant, especially for the longer gate operating times required at high $\Delta$ and high $t_c^0$. \label{fig:Fvtc}}
	\end{figure*}
It is worth emphasizing that in all cases the minimum achievable infidelity is only mildly dependent on $t_c$ provided $t_c>B_z$.  We might anticipate that the critical determinants of the optimal operating parameters will be related to low-frequency noise. We summarize results including charge and hyperfine noise in the low-frequency limit in Fig.~\ref{fig:Fvtc}, where we again plot the average infidelity with respect to $\Delta$.  Low-frequency charge noise only weakly impairs fidelity and shows a mild $\Delta$ dependence generally. However, the impact of hyperfine noise is dramatic, showing a strong dependence on both $\Delta$ and $t_c$ due to the longer gate operation times as both of these increase.  The effect is strong enough to negate any advantages associated with operating in the spin-like regime, indicating the importance of optimizing parameters that minimize the operating time of the gate.  Additionally, upon comparing the $t_c^0$ = 25 $\mu$eV and $t_c^0$ = 40 $\mu$eV cases it is apparent that significant reductions in hyperfine noise are required to shift optimal operation to higher $t_c^0$ and meaningfully improve the fidelity.   

\begin{figure}
	\centering
	\includegraphics[width=\columnwidth]{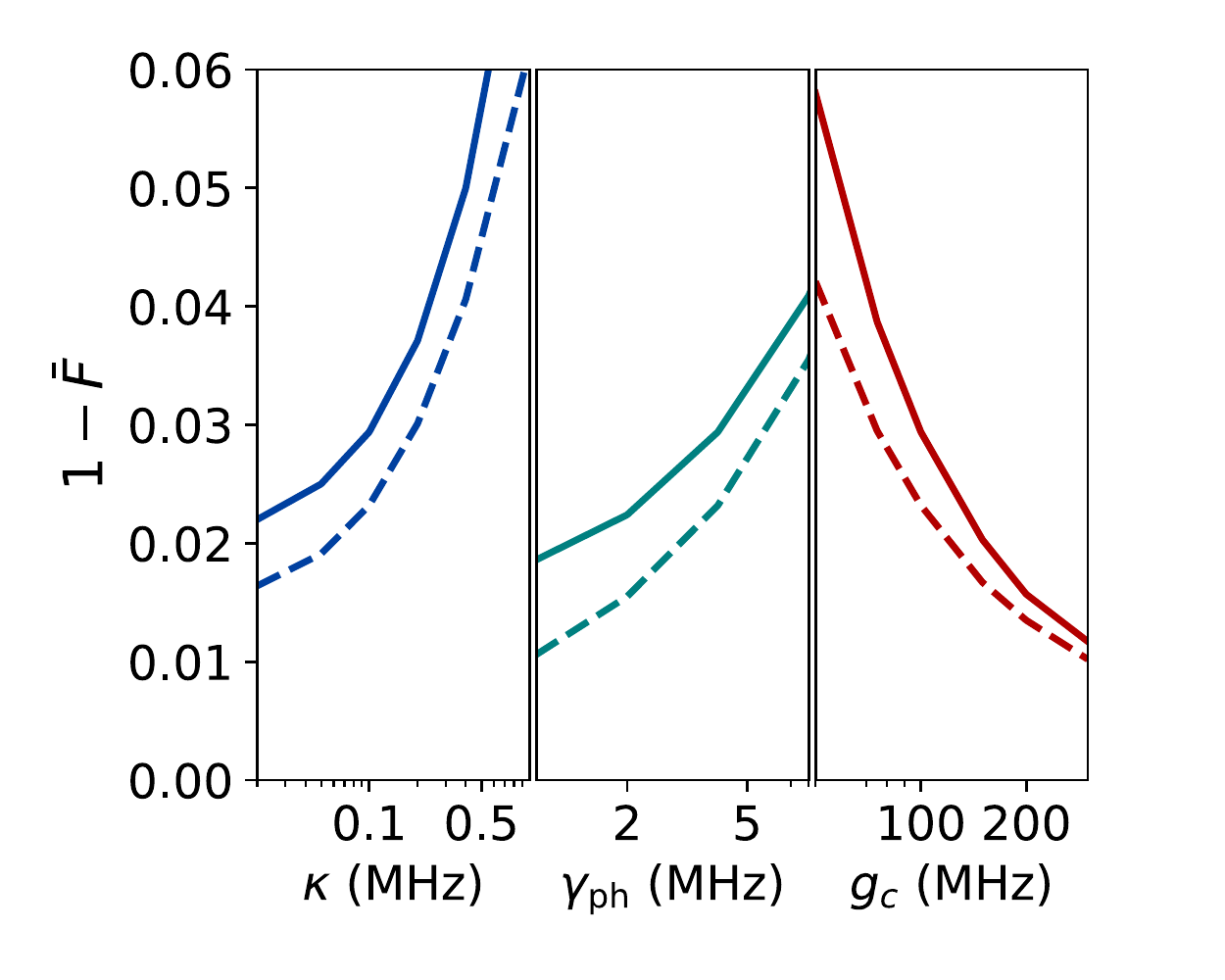}
	\caption{Dependence of the minimum infidelity with (solid) and without (dashed) low frequency noise on the parameters $\kappa$, $\gamma_{\rm ph}$, and $g_c$ from the parameter space point $\kappa=0.1$MHz, $\gamma_{\rm ph}=4$MHz, and $g_c=100$MHz.  In all cases $t_c^0=25\mu$eV,$\gamma_{\rm ch}=\gamma_{\rm ph}$, and $v=$2meV.}\label{fig:optima}
\end{figure}

In Fig.~\ref{fig:optima} we show the optimal infidelity as a function of several parameters.  In the first panel we show the effect of reductions in cavity loss rate $\kappa$.  As expected, increasing cavity quality allows for operation in regimes with reduced gate times and significant improvements to fidelity at smaller values of $\kappa$.  However, once $\kappa$ is reduced to $\lesssim 0.1$MHz cavity loss is no longer the dominant fidelity limiter; Landau-Zener behavior becomes the primary obstacle to reductions in gate time and noise is dominated by charge noise.  The second panel shows the impact of changes in charge noise; it is evident that in the regime under consideration reductions in charge noise offer modest improvements to fidelity.  The third panel shows the impact of increases to the cavity coupling $g_c$.  Increases to $g_c$ allow for smaller spin-mixing angles to achieve a given $g_s$, and thus reduced exposure to charge noise. As a result, fidelity gains from increasing cavity coupling are substantial; doubling $g_c$ nearly halves the infidelity over the whole range shown.  

\section{Effects of Si valley physics}

\begin{figure}
	\centering
	\includegraphics[width=\columnwidth]{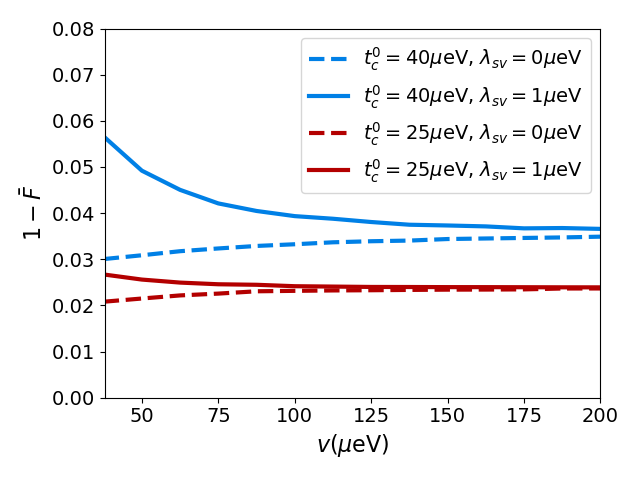}
	\caption{ Simulated infidelity without the inclusion of low-frequency noise for different $t_c^0$ allowing the valley splitting $v$ to vary. The infidelities shown are assuming optimization of the gate duration for $\Delta=20$ MHz ($t_c^0$ = 25 $\mu$eV) and $\Delta$ = 8 MHz ($t_c^0$ = 40 $\mu$eV). Other parameters are $g_c$ = 100 MHz, $\gamma_{\rm ph}=\gamma_{\rm ch}$ = 4 MHz, $\kappa$ = 0.1 MHz for all cases. As can be seen, valley splitting exerts significant influence only when spin-vally coupling is present and $v$ is less than about $2t_c^0$.\label{fig:FvVS}}
\end{figure}

We now consider the influence of the valley degree of freedom on the iSWAP fidelity. In Fig.~\ref{fig:FvVS} we plot the difference in infidelity from the infinitely large valley splitting case against valley splitting $v$ (we assume both DQDs have the same splitting).  We show the case for parameter set $g_c$ = 100 MHz, $\gamma_{\rm ph}=\gamma_{\rm ch}$ =4 MHz, $\kappa$ = 0.1 MHz for different $t_c^{0}$; we exclude low-frequency noise due to computational expense and since we are interested in a differential effect.  One might expect low lying valley states to lead to additional loss and reduced fidelity; the extent of this impairment is strongly dependent on $t_c^0$.  While for $t_c^0$ = 40 $\mu$eV infidelity becomes significant at around $v<125$ $\mu$eV, for $t_c^0$ = 25 $\mu$eV, the impact is negligible until $v<50$ $\mu$eV.  This is a consequence of spin-valley coupling, as seen in the infidelity when spin-valley coupling is set to zero and infidelity actually falls slightly as valley splitting is decreased.  This is easy to understand if we consider the DQD qubit states.  From Fig.~\ref{fig:single_DQD} we note that while the two lowest states can acquire significant valley character, they acquire mostly the same valley character; neither dephasing between valleys nor relaxation from upper valley states will exert an effect.  Spin-valley coupling, however, allows these terms to influence the dynamics by introducing an asymmetry in the valley character of the two qubit states, creating a channel for valley dephasing and relaxation.  This asymmetry is small however, and manifests strongly only over long gate operating times imposed by larger $t_c^0$.

\section{Conclusion}
\label{sec:Conclusion}

Full simulations of a spin-cavity-spin iSWAP gate reveal performance that depends critically on the choice of operating parameters in a way that is characterized by multiple tradeoffs.  Fidelity is of course limited by noise, but the management of the noise is nontrivial, especially when valley splitting is low and spin-valley coupling provides additional noise channels. High-frequency charge, phonon, and hyperfine noise all favor shorter gate operation times, suggesting that one should attempt to operate in the regime where qubit-cavity coupling is higher and the cavity is closer to resonant with the qubits.  However, this coupling also results in stronger population of the cavity during operation, makes the system prone to Landau-Zener transitions to states outside the subspace of the qubits and  increases loss due to emission from the cavity. Both of these factors  -- loss due to photon coupling for shorter gate operating times and loss due to noise, especially hyperfine, at longer gate operating times -- have strong dependence on the main operating parameters $t_c^0$ and $\Delta$; these must be chosen carefully based on the characteristics of the device.  

These results offer clear guidance in terms of device improvement. The involvement of excited valley states is not an overarching concern with regards to additional noise or the possibility of transitions to unwanted states in the preferred coupling regimes, as moderately low valley splittings are tolerable in the suggested operating regime. Phonon relaxation and charge noise strongly impair performance but are challenging to reduce significantly. While previously ignored hyperfine noise  has a significant impact, it can  in general can be reduced through the use of isotopically pure silicon and otherwise mitigated by choice of operating regime.  On the other hand, cavity loss can be substantial in the optimal operating regime and improvements in cavity quality are much more feasible; however, returns to reductions in cavity loss diminish once non-adiabaticity becomes the dominant limiter to shorter operating times. Finally, enhancing $g_c$ is most critical, as it allows for faster gate operations without exposure to greater charge and phonon noise (as a result of taking on more charge character); doubling $g_c$ essentially halves the achievable infidelity.  It will be essential to take advantage of such improvements to operate close to the near-resonant regime where faster gate times are possible.

\section{Acknowledgements}
Sandia National Laboratories is a multi-mission laboratory managed and operated by National Technology and Engineering Solutions of Sandia, LLC., a wholly owned subsidiary of Honeywell International, Inc., for the U.S. Department of Energy's National Nuclear Security Administration under contract DE-NA-0003525. This paper describes objective technical results and analysis. Any subjective views or opinions that might be expressed in the paper do not necessarily represent the views of the U.S. Department of Energy or the United States Government. JRP acknowledges support from Army Research Office grant W911NF-15-1-0149.

\bibliography{bibliography}
\bibliographystyle{unsrt}
\cleardoublepage

\end{document}